\documentclass[12pt]{article}
\usepackage{epsfig}
\usepackage{graphicx}
\usepackage{citesort}
\usepackage{amsmath,amssymb,amsfonts}

\usepackage[dvips,a4paper,textwidth=15.0cm,textheight=21.0cm]{geometry}
\newcommand{\beq}{\begin{equation}}
\newcommand{\eeq}{\end{equation}}
\newcommand{\bea}{\begin{eqnarray}}
\newcommand{\eea}{\end{eqnarray}}

\newcommand\Tr{\operatorname{Tr}}

\renewcommand\Re{\operatorname{Re}}

\def\dd{{\mathrm{d}}}

\begin{document}
\title{\bf Polyakov loop, diquarks and the two-flavour phase diagram \footnote{Work supported in part by BMBF, GSI and INFN}}
\author{
S. R\"o{\ss}ner$^{a}$, C. Ratti$^{b}$ and W. Weise$^{a}$\\
\\{\small $^a$ Physik-Department, Technische Universit\"at M\"unchen, D-85747 
Garching, Germany}\\
{\small $^b$ ECT$^*$, I-38050 Villazzano (Trento) Italy and INFN, Gruppo Collegato di Trento, }\\
{\small via Sommarive, I-38050 Povo (Trento) Italy}
}
\date{February 9, 2007} 
\maketitle
\begin{abstract}
An updated version of the PNJL model is used to study the thermodynamics  of
$N_f = 2$ quark flavours interacting through chiral four-point couplings and propagating in a homogeneous Polyakov loop background. Previous PNJL calculations are extended by introducing explicit diquark degrees of freedom and an improved effective potential for the Polyakov loop field. The mean field equations are treated under the aspect of accommodating group theoretical constraints and issues arising from the fermion sign problem. The input is fixed exclusively by selected pure-gauge lattice QCD results and by pion properties in vacuum. The resulting $(T, \mu)$ phase diagram is studied with special emphasis on
the critical point, its dependence on the quark mass and on Polyakov loop dynamics. We present successful comparisons with lattice QCD thermodynamics expanded to finite chemical potential $\mu$.

\end{abstract}
\section{Introduction}

Reconstructing the phase diagram and thermodynamics of QCD in terms of field theoretical quasiparticle models is an effort worth pursuing in order to interpret lattice QCD 
results \cite{phil,katz,Allton1,Allton:2005gk,Boyd,Kaczmarek:2002mc,Kaczmarek:2005} and extrapolate into
regions not accessible by lattice computations. A promising ansatz of this sort is the PNJL model
\cite{Fukushima:2003fw,Meisinger:1995,ratti,Ratti:2006wg}, a synthesis of Polyakov loop dynamics with the Nambu $\&$ Jona-Lasinio model, combining the two principal non-perturbative features of low-energy QCD: confinement and spontaneous chiral symmetry breaking.  This paper extends our previous PNJL calculations \cite{ratti,Ratti:2006wg} in several directions. First, diquark degrees of freedom are explicitly included. Diquark condensation at large quark chemical potential is explored in the presence of a Polyakov loop background. Secondly, in comparison with our previous work, the effective potential which controls the thermodynamics of the Polyakov loop field is improved such that group theoretical constraints are implemented, following  Ref.\cite{Fukushima:2003fw}. 
Issues of the mean field approximation in the context of the fermion sign problem are discussed in comparison with previous work.

The aim of the present paper is to investigate the phase diagram resulting from this approach. Of special interest is the location of the critical point, its dependence on the quark mass and the role of the Polyakov loop as indicator of the deconfinement transition. Predictions for the leading coefficients in a Taylor expansion of the pressure in powers of the quark chemical potential will turn out to be remarkably successful in comparison with corresponding lattice QCD results.

\section{The PNJL model}

The two-flavour PNJL model (now including diquark degrees of freedom) is specified by the Euclidean
action
\beq 
{\cal S}_E(\psi, \psi^\dagger, \phi)= \int _0^{\beta=1/T} d\tau\int d^3x \left[\psi^\dagger\,\partial_\tau\,\psi + {\cal H}(\psi, \psi^\dagger, \phi)\right] + \delta{\cal S}_E(\phi,T)
\label{action}
\eeq 
with the fermionic Hamiltonian density \footnote{$\vec{\alpha} = \gamma_0\,\vec{\gamma}$ and $\gamma_4 = i\gamma_0$ in terms of the standard Dirac $\gamma$ matrices.}:
\beq
{\cal H} = -i\psi^\dagger\,(\vec{\alpha}\cdot \vec{\nabla}+\gamma_4\,m_0 -\phi)\,\psi + {\cal V}(\psi, \psi^\dagger)~,
\label{H}
\eeq
where $\psi$ is the $N_f=2$ doublet quark field and $m_0 = diag(m_u,m_d)$ is the quark mass matrix. The quarks move in a background color gauge field $\phi \equiv A_4 = iA_0$, where $A_0 = \delta_{\mu 0}\,g{\cal A}^\mu_a\,t^a$ with the $SU(3)_c$ gauge fields ${\cal A}^\mu_a$ and the generators $t^a = \lambda^a/2$. The matrix valued, constant field $\phi$ relates to the (traced) Polyakov loop as follows:
\beq
\Phi=\frac{1}{N_c}\mathrm{Tr}\left[\mathcal{P}\exp\left(i\int_{0}^{\beta}
d\tau A_4\right)\right]=\frac{1}{ 3}\mathrm{Tr}\,e^{i\phi/T}~,\label{eqn:polyakovloop}
\eeq
In a convenient gauge (the so-called Polyakov gauge), the matrix $\phi$ is given a diagonal representation
\beq
\phi = \phi_3\,\lambda_3 +  \phi_8\,\lambda_8~,
\eeq
which leaves only two independent variables, $\phi_3$ and $\phi_8$. The piece $\delta{\cal S}_E = \frac{V}{ T}{\cal U}$ of the action (\ref{action}) controls the thermodynamics of the Polyakov loop. It will be specified later in terms of the effective potential, ${\cal U}(\Phi,T)$, determined such that the thermodynamics of pure gauge lattice QCD is reproduced for $T$ up to about twice the critical temperature for deconfinement.
At much higher temperatures where transverse gluons begin to dominate, the PNJL model is not supposed to be applicable. 

The interaction ${\cal V}$ in Eq.~(\ref{H}) includes chiral $SU(2)\times SU(2)$ invariant four-point couplings of the quarks acting in
pseudoscalar-isovector/scalar-isoscalar quark-antiquark and scalar diquark channels:  
\bea
\mathcal{V}= -\frac
{G}{2}\left[\left(\bar{\psi}\psi\right)^2+\left(\bar{\psi}\,i\gamma_5
\vec{\tau}\,\psi
\right)^2\right]
- \frac{H}{2}\left[\left(\bar{\psi}\,{\cal C}\gamma_5\tau_2\lambda_2
\,\bar{\psi}^{T}\right)\left(\psi^{T}\gamma_5\tau_2\lambda_2 {\cal C}
\,\psi\right)\right]~,
\label{V}
\eea
where ${\cal C}$ is the charge conjugation operator. We can think of Eq.~(\ref{V}) as a subset in the chain of terms generated by Fierz-transforming a local color current-current interaction
between quarks, 
\bea
{\cal L}_{int} = - G_c(\bar{\psi}\gamma_\mu t^a\psi)(\bar{\psi}\gamma^\mu t^a\psi)~.
\nonumber
\eea
In this case the coupling strengths in the quark-antiquark and diquark sectors are related by $G = \frac{4}{3}H$, the choice
we adopt. The minimal ansatz (\ref{V}) for ${\cal V}$ is motivated by the fact that spontaneous chiral symmetry breaking is driven by the first term while the second term induces diquark condensation at sufficiently large chemical potential of the quarks. Additional pieces representing vector and axialvector $q\bar{q}$ excitations as well as color-octet diquark and  $q\bar{q}$ modes are omitted here. We have checked that their effects are not important in the present context.  

The NJL part of the model involves three parameters: the quark mass which we take equal for $u$- and $d$-quarks, the coupling strength $G$ and a three-momentum cutoff $\Lambda$. We take those from Ref.\cite{ratti}:
\bea
m_{u,d} = 5.5~\mathrm{MeV}~,~~G = \frac{4}{3}H = 10.1~\mathrm{GeV}^{-2}~,~~\Lambda = 0.65~\mathrm{GeV}~,
\nonumber
\eea 
which were fixed to reproduce the pion mass and decay constant in vacuum and the chiral condensate as $m_\pi =$ 139.3 MeV, $f_\pi =$ 92.3 MeV and $\langle\bar{\psi}_u\psi_u\rangle = - (251$ MeV)$^3$.

The effective potential $\mathcal{U}(\Phi,T)$ which controls the dynamics of the Polyakov loop has the following properties. It must satisfy the $Z(3)$ center symmetry of the pure gauge QCD Lagrangian.
In the low-temperature (confinement) phase $\mathcal{U}(\Phi, T)$ has an absolute minimum at $\Phi = 0$. 
Above the critical temperature for deconfinement ($T_0\simeq$ 270 MeV 
according to pure gauge lattice QCD results) the $Z(3)$ symmetry is spontaneously broken and the minimum of  $\mathcal{U}(\Phi, T)$ is shifted to a finite value of $\Phi$. In the 
limit $T\rightarrow\infty$ we have $\Phi\rightarrow~1$. 

In our previous Ref.\cite{ratti} the simplest possible polynomial form was chosen
for ${\cal U}$. In the present work an improved expression, guided by Ref.\cite{Fukushima:2003fw}, replaces the higher order polynomial terms in $\Phi, \Phi^*$  
by the logarithm of $J(\Phi)$,  the Jacobi determinant which results from integrating out six non-diagonal Lie algebra directions while keeping the two diagonal ones, $\phi_{3,8}$, to represent $\Phi$.
This suggests the following ansatz for $\cal{U}$:
\beq
\frac{{\cal U}(\Phi,T )}{T^4}=-\frac{1}{2}a(T)\,\Phi^*\Phi
+b(T)\,\ln\left[1-6\,\Phi^*\Phi+4\left({\Phi^*}^3+\Phi^3\right)
-3\left(\Phi^*\Phi\right)^2\right]
\label{u1}
\eeq
with 
\beq
a(T)=a_0+a_1\left(\frac{T_0}{T}\right)
+a_2\left(\frac{T_0}{T}\right)^2,~~~~~~b(T)=b_3\left(\frac{T_0}{T}
\right)^3.
\label{u2}
\eeq
With its logarithmic divergence
as $\Phi,\Phi^*\rightarrow 1$, this ansatz automatically limits the Polyakov loop $\Phi$ to be always smaller than 1, reaching this value
asymptotically only as $T\rightarrow\infty$.
Following the procedure as in~\cite{ratti}, a precision fit of the parameters 
$a_i$ and $b_3$ is performed in order to reproduce lattice data for
pure gauge QCD thermodynamics and for the behaviour of the Polyakov loop
as a function of temperature. 
The critical temperature $T_0$ for deconfinement in the pure gauge sector is
fixed at 270 MeV in agreement with lattice results.

\begin{figure}
\begin{center}
\parbox{.48\textwidth}{ 
\includegraphics*[height=170pt, clip=true, trim=18 4 2 5 ] {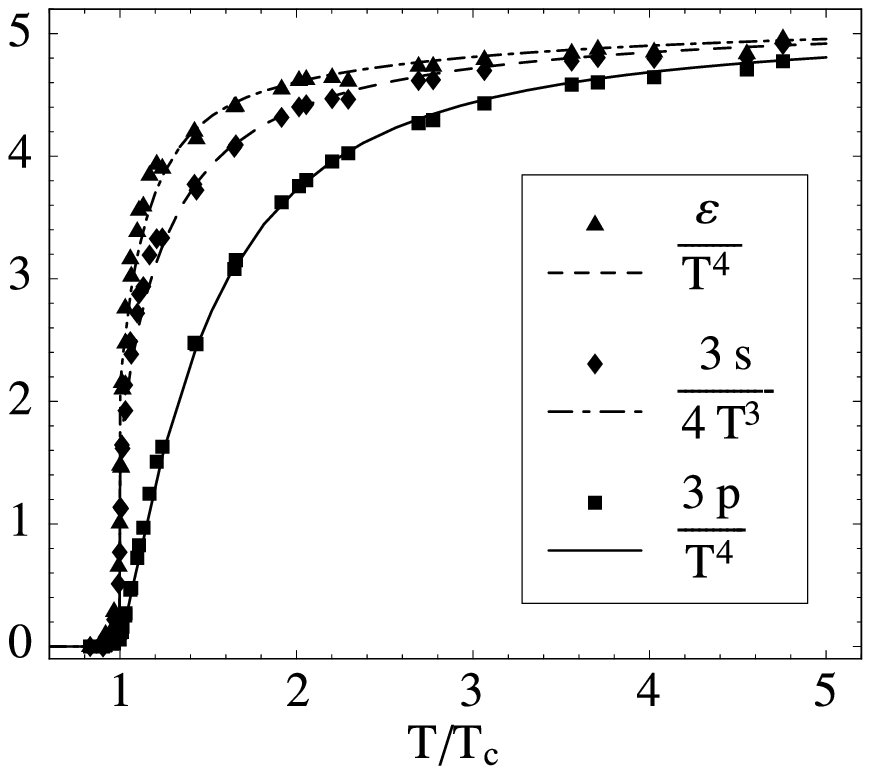} }
\hfill
\parbox{.51\textwidth}{\hfill  
\includegraphics*[height=170pt, clip=true, trim=25 0 40 0 ] {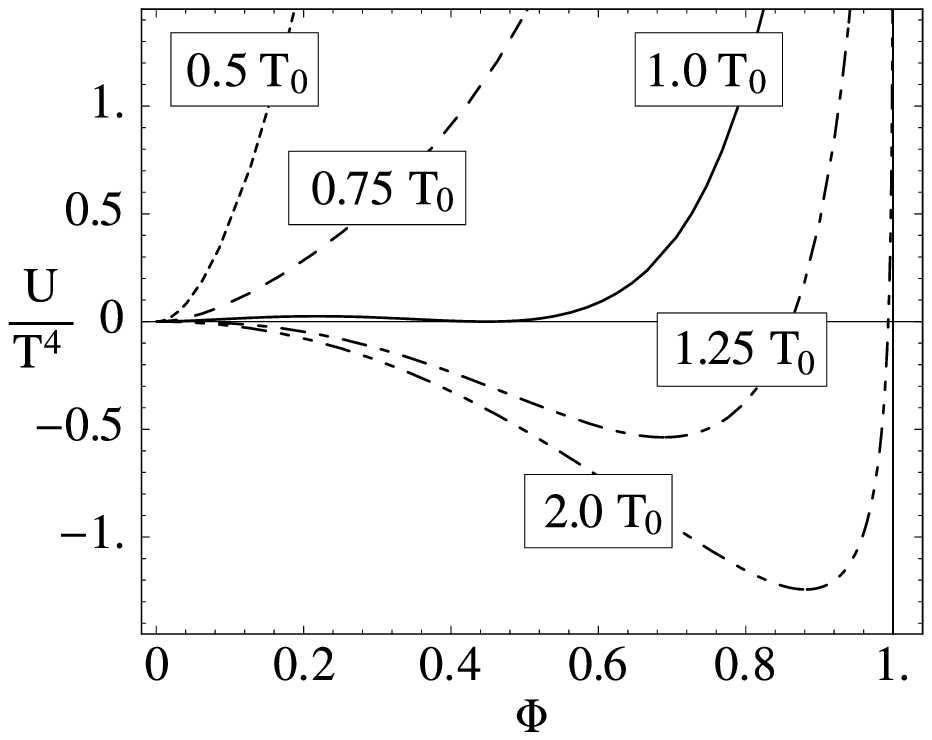} }
\caption{
\footnotesize
Left: Fit to scaled pressure, entropy density and energy density as functions
of the temperature in the pure gauge sector, compared to the corresponding 
lattice data taken from Ref.~\cite{Boyd}. Right: Resulting effective potential (\ref{u1}) that drives spontaneous Z(3) symmetry breakdown at $T=T_0$.
\label{fig:fitted_potential}}
\end{center}
\end{figure}

The results of this combined fit are shown in Fig.~\ref{fig:fitted_potential} and the dotted line  of Fig.~\ref{fig:polyakovloop}. The corresponding parameters are 
\bea
a_0 = 3.51~,~~a_1 = -2.47~,~~a_2 = 15.2~,~~b_3 = -1.75~.
\nonumber
\eea
The fit was constrained by demanding that the Stefan-Boltzmann limit is reached within the model at $T\to \infty$ and by enforcing a first-order phase transition at $T_0$. 
The first constraint determines $a_0 = \frac{16\pi^2}{45} \approx 3.51$. The second constraint fixes $b_3 = -0.108\,(a_0 + a_1 + a_2)$. The two remaining parameters $a_1$ and $a_2$ were determined using a mean square fit. In this fit the Polyakov loop data set \cite{Kaczmarek:2002mc} was given a stronger weight than the pressure, energy density and entropy \cite{Boyd}. This was done in order to counterbalance the smaller number of Polyakov loop data points against the larger number of data sets for pressure, energy density and entropy. The resulting uncertainties are estimated to be about $6\,\%$ for $a_1$ and less than $3\,\%$ for $a_2$. These independent errors propagate to $b_3$ giving an uncertainty of about $2\,\%$.

Next, the PNJL action is bosonized and rewritten in 
terms of the auxiliary scalar and pseudoscalar fields ($\sigma,\vec{\pi}$), and diquark and antidiquark fields ($\Delta,\Delta^*$). The thermodynamic potential of the model is evaluated as follows:
\bea
\Omega=\mathcal{U}\left(\Phi,T\right)-\frac{T}{2}\sum_n\int\frac{d^3p}
{\left(2\pi\right)^3}\mathrm{Tr}\ln\left[\beta \tilde{S}^{-1}\left(i\omega_n,\vec{p}\,\right)
\right]+\frac{\sigma^2}{2G}+\frac{\Delta^*\Delta}{2H}~,
\label{omega1}
\eea
where the sum is taken over Matsubara frequencies $\omega_n = (2n+1)\pi T$. The inverse Nambu-Gor'kov propagator including diquarks is:
\beq
\tilde{S}^{-1}\left(i\omega_n,\vec{p}\,\right)=\left({{\begin{array}{ccc} 
i\gamma_0\,\omega_n-\vec{\gamma}\cdot\vec{p}-
m+\gamma_0\left(\mu - i \phi\right)& \Delta\gamma_5\tau_2\lambda_2\\
-\Delta^*\gamma_5\tau_2\lambda_2& 
i\gamma_0\, \omega_n-\vec{\gamma}\cdot\vec{p}-
m-\gamma_0\left(\mu - i \phi\right)
\end{array}}}\right)~.
\label{propmu}
\eeq
Just as in the standard NJL model, quarks develop a dynamical (constituent) mass through
their interaction with the chiral condensate:
\beq
m=m_0-\langle\sigma\rangle=m_0-G\langle\bar{\psi}\psi\rangle.
\label{mass}
\eeq
With the input parameters previously specified one finds $m =$ 325 MeV at $T = 0$. 

Note that introducing diquarks (and anti-diquarks) as explicit degrees of freedom implies off-diagonal
pieces in the inverse propagator (\ref{propmu}). As a consequence, the traced Polyakov loop field $\Phi$ and its conjugate $\Phi^*$ can no longer be factored out when performing the $\Tr \ln = \ln \det$ operation in the thermodynamic potential (\ref{omega1}), unlike the simpler case treated in our previous Ref.~\cite{ratti}. The explicit evaluation of energy eigenvalues now involves $\phi_3$ and $\phi_8$ as independent fields. The final result for $\Omega$ is then given as:
\bea
\Omega & = &{\cal U}\left(\Phi,T\right)+\frac{\sigma^2}{2G}+
\frac{\Delta^*\Delta}{2H}
\nonumber\\
 & - & 2N_f\int\frac{d^3p}{\left(2\pi\right)^3}\sum_j \left\{
T\ln\left[1+e^{-E_j/T}\right]+\frac12 \Delta E_j
\right\}~.
\label{omega}
\eea
The difference $\Omega(T)-\Omega(T=0)$ is to be used in the actual calculations. 
The quasi-particle energies $E_j$, denoted by indices $j$ running from $1$ to $6$, have the following explicit expressions with $\varepsilon(\vec{p}\,) = \sqrt{\vec{p}\,^2+m^2}$:
\bea
E_{1,2}&=&\varepsilon(\vec{p}\,)\mp\tilde{\mu}_b~,
\nonumber\\
E_{3,4}&=&\sqrt{(\varepsilon(\vec{p}\,)+\tilde{\mu}_r)^2+|\Delta|^2}\mp i\,\phi_3~,
\nonumber\\
E_{5,6}&=&\sqrt{(\varepsilon(\vec{p}\,)-\tilde{\mu}_r)^2+|\Delta|^2}\mp i\,\phi_3~,
\eea
with
\bea
\tilde{\mu}_b=\mu+2i\,\frac{\phi_8}{\sqrt{3}}~,~~~~~~~~~~~
\tilde{\mu}_r=\mu-i\,\frac{\phi_8}{\sqrt{3}}~.
\eea
In Eq.(\ref{omega}), $\Delta E_j = E_j - \varepsilon \pm \mu$ is the difference of the quasiparticle energy $E_j$ and the energy of free fermions (quarks), where the upper sign applies for fermions and the lower sign for antifermions. It is understood that for three-momenta $|\vec{p}\,|$ above the cutoff $\Lambda$ where NJL interactions are "turned off", the quantities $\sigma$ and $\Delta, \Delta^*$ are set to zero.

The thermodynamical potential (\ref{omega}) involves the bosonic field variables $\sigma$, $\Delta$, $\phi_3$ and $\phi_8$.
In the mean field theory the integration over all field configurations $\left\lbrace \sigma,\,\Delta,\,\phi_3,\,\phi_8\right\rbrace$ in the calculation of the partition function is approximated by a single field configuration, $\left\lbrace \sigma,\,\Delta,\,\phi_3,\,\phi_8\right\rbrace_{\mathrm{m.f.}}$.
For a purely real action the optimal mean field configuration is the one which determines a minimum of $\Omega = \Omega(\sigma,\,\Delta,\,\phi_3,\,\phi_8)$.
The necessary condition for this is
\beq
\frac{\partial\,\Omega}{\partial\left( \sigma, \Delta, \phi_3, \phi_8\right)}=0 . \nonumber
\eeq
Here the action is treated in analogy with a Landau effective action which identifies the fields with approximate order parameters. In case of the PNJL model, however, the thermodynamical potential $\Omega$ is complex in the presence of the Polyakov loop background and at non-zero chemical potential $\mu$.
A minimization of a complex function as such is void of meaning.
This is the so-called fermion sign problem in the present context.
However, even for a complex Euclidean action $\mathcal{S}_{\mathrm{E}}$, one can still search for the configuration with the largest weight in the path integral and refer to this as the mean field configuration.
An analysis of the phase and the absolute value of the weight $e^{-\mathcal{S}_{\mathrm{E}}}$ immediately shows that this mean field configuration maximizes $\vert e^{-\mathcal{S}_{\mathrm{E}}} \vert$ and consequently minimizes $\Re \Omega$.
The mean field equations derived from $\Omega$ are
\beq
\frac{\partial\,\Re\,\Omega}{\partial \left( \sigma, \Delta, \phi_3, \phi_8\right)}=0,
\label{mf}
\eeq
the condition we adopt.

In previous publications \cite{ratti,Ghosh:2006qh,Zhang:2006gu}, the last two mean field equations, $\frac{\partial \Omega}{\partial \phi_3} = \frac{\partial \Omega}{\partial \phi_8} = 0$, were replaced by $\frac{\partial \Omega}{\partial \Phi} = \frac{\partial \Omega}{\partial \Phi^*} = 0$.
These relations are in principle equivalent, given that $\Phi=\Phi(\phi_3,\,\phi_8)$ and $\Phi^*=\Phi^*(\phi_3,\,\phi_8)$ as implied by Eq. (\ref{eqn:polyakovloop}). The constraint under which such a change of variables can be done is that the temporal gauge fields remain real quantities: $\phi_3,\,\phi_8\in \mathbb{R}$.

Abandoning this constraint would introduce different chemical potentials for quarks of different colors.\footnote{This can formally be seen when using the simple prescription $\mu \longrightarrow \mu-i \left(\phi_3\,\lambda_3\,T + \phi_8\,\lambda_8\,T \right)$ to do the transition from the NJL- to the PNJL-Nambu-Gor'kov propagator.}
Using equation~(\ref{eqn:polyakovloop}) it can easily be derived that in the case where $\phi_3\in \mathbb{R}$ and $\phi_8 \in \mathbb{R}$, $\Phi$ and $\Phi^*$ genuinely have to be the complex conjugate of each other.

The (thermal) expectation values $\langle \Phi \rangle$ and $\langle \Phi^* \rangle$ of the conjugate Polyakov loop fields must be real quantities as argued in \cite{dpz05}.
This applies to the PNJL model as well, in the sense that the mean field action $\mathcal{S}_{\mathrm{MF}}$ changes into its complex conjugate under charge conjugation. Enforcing $\langle \Phi \rangle \in \mathbb{R}$ and $\langle \Phi^* \rangle \in \mathbb{R}$ means $ \Phi = \Phi^*$ for the mean field configurations that satisfy Eq.~(\ref{mf}).
With the constraint of $\phi_3$ and $\phi_8$ being real, this implies $\phi_8=0$ leaving only $\phi_3$ as an independent variable.

In previous work \cite{ratti,Ghosh:2006qh,Zhang:2006gu} $\langle \Phi \rangle$ and $\langle \Phi^* \rangle$ have been treated as independent real quantities in the minimization of $\Omega$.
This procedure, without the constraints imposed by $\phi_3,\,\phi_8 \in \mathbb{R}$, tends to overestimate the difference between $\langle \Phi \rangle$ and $\langle \Phi^* \rangle$.
While this unphysical feature has only marginal consequences for global properties such as the pressure, it does have a visible influence on more detailed quantities as discussed in section~\ref{sec:finitemu}

Within the present mean field context defined by Eq.~(\ref{mf}), fluctuations {\cal beyond} mean field are at the origin of $\langle\Phi\rangle\neq\langle\Phi^*\rangle$ for $\mu\neq 0$.
This paper deals with self-consistent solutions and predictions of the mean-field equations ($\ref{mf}$). While further extensions including quantum fluctuations~\cite{simon1} will be subject of a forthcoming publication~\cite{simon2}, we can already anticipate one of the results, namely that the effects of fluctuations, leading to $\langle\Phi\rangle\neq\langle\Phi^*\rangle$ at finite chemical potential, turn out not to be of major qualitative importance in determining the phase diagram. This forthcoming publication will present a way how to deal with the fermion sign problem in the context of the PNJL model. Discussions of the fermion sign problem in other models can be found in \cite{dpz05,Fukushima:2006uv}.

\section{Results}

\begin{figure}
\begin{center}
\includegraphics*[width=.52\textwidth]{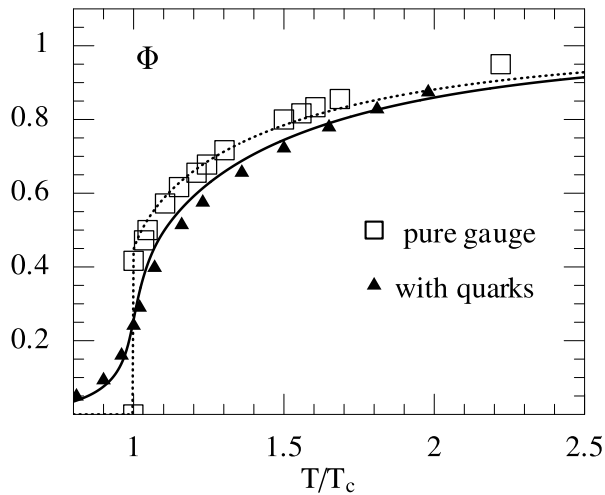}
\caption{
\footnotesize
Using the fit of the Polyakov loop (dotted line) to lattice results taken from Ref.~\cite{Kaczmarek:2002mc} in the pure gauge sector (empty symbols), the PNJL model predicts the Polyakov loop behaviour as a function of temperature in the presence of dynamical quarks (solid line). This prediction is compared to lattice data in two flavours (full symbols) taken from Ref.~\cite{Kaczmarek:2005}.
\label{fig:polyakovloop}}
\end{center}
\end{figure}

\begin{figure}
\begin{center}
\includegraphics*[width=.5\textwidth]{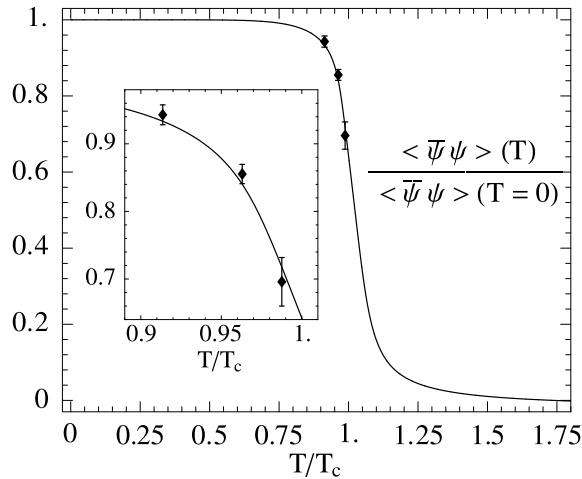}
\caption{
\footnotesize
The spontaneous chiral symmetry breaking mechanism of the PNJL model generates a temperature dependent chiral condensate $\left\langle \bar{\psi}\psi \right\rangle$ (solid line), which is compared here to lattice QCD results in two flavours shown in Ref.~\cite{Boyd:1995cw}.
\label{fig:chrialcondensate}}
\end{center}
\end{figure}

Solution of the mean-field equations~(\ref{mf}) yields the chiral condensate, $\langle\bar{\psi}\psi\rangle = \sigma/G$, the color-antitriplet diquark condensate, $\Delta$, and the Polyakov loop exponent $\phi_3$ as functions of $T, \mu$. The resulting prediction for the traced Polyakov loop $\Phi$ at $\mu=0$ (where the diquark condendate vanishes, $\Delta=0$) is shown in Fig.~\ref{fig:polyakovloop} (continuous line)
in comparison with the corresponding lattice data taken from 
Ref.~\cite{Kaczmarek:2005} (full symbols). 
The agreement is quite remarkable. The improvement in comparison to previous calculations is primarily due to the improved ansatz for the Polyakov loop potential.
In the presence of quarks, the deconfinement transition
is no longer first-order as in pure gauge QCD. It becomes a 
smooth crossover when quarks couple to the Polyakov loop field. 
The critical temperature for deconfinement is now decreased from
$270\,\mathrm{MeV}$ to a smaller value\footnote{Note that the critical temperature in full lattice QCD reported in \cite{Kaczmarek:2005} is $T_c = 202\,\mathrm{MeV}$.} around $215\,\mathrm{MeV}$ (not evident from 
Fig.~\ref{fig:polyakovloop} where the results are plotted as functions of $T/T_c$). In Fig.~\ref{fig:chrialcondensate} we show in addition the predicted temperature dependence of the two-flavour chiral condensate $\langle\bar{\psi}\psi\rangle$ in comparison with lattice data \cite{Boyd:1995cw}.

\subsection{Finite chemical potential}\label{sec:finitemu}
\begin{figure}
\begin{minipage}[t]{.48\textwidth}
\hfill 
\includegraphics*[width=.93\textwidth]{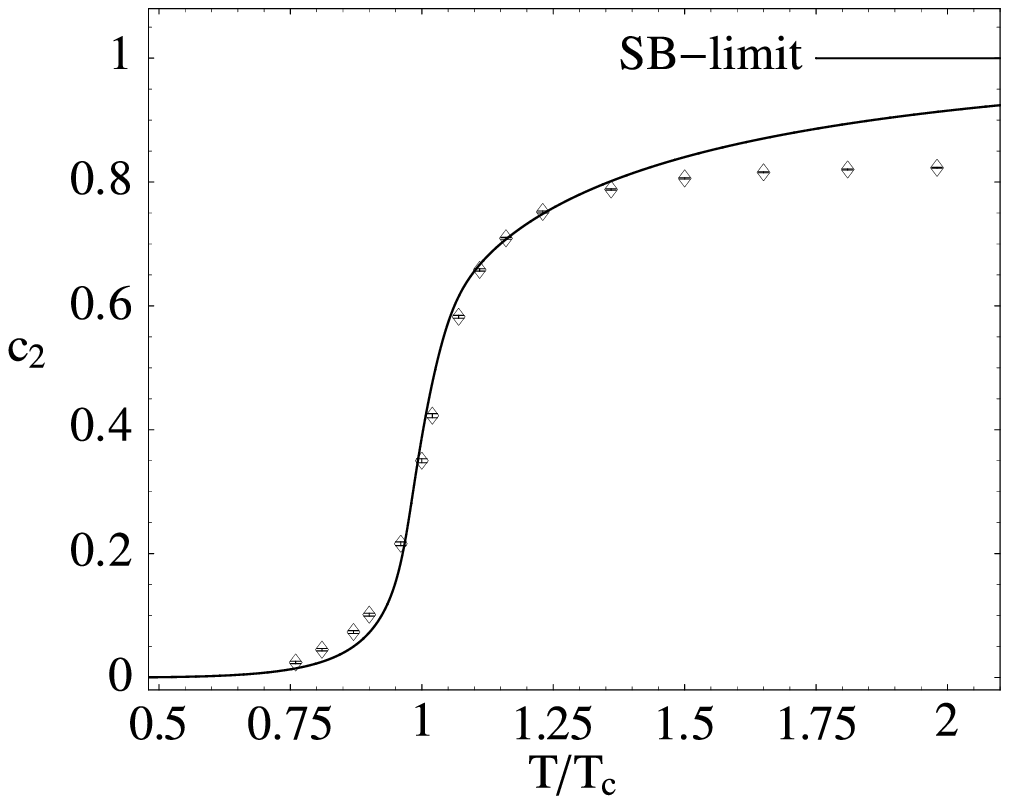}
\end{minipage}
\hfill
\begin{minipage}[t]{.48\textwidth}
\hfill 
\includegraphics*[width=.95\textwidth]{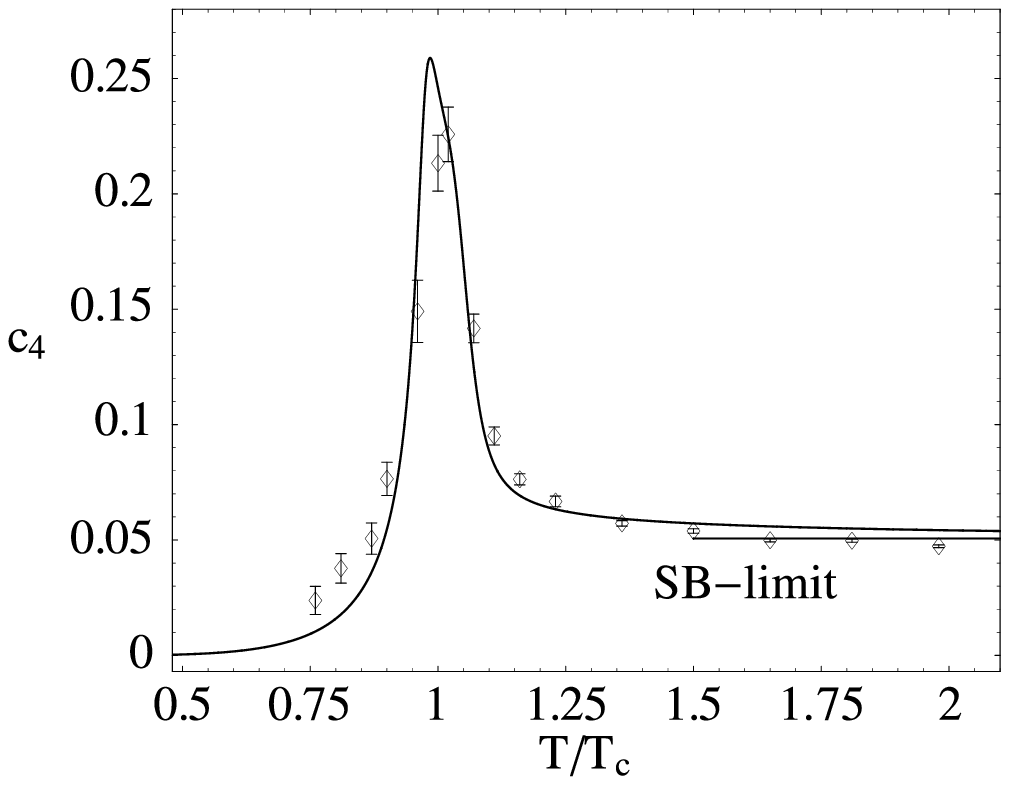}
\end{minipage}
\\
\begin{minipage}[t]{.48\textwidth}
\includegraphics*[width=\textwidth]{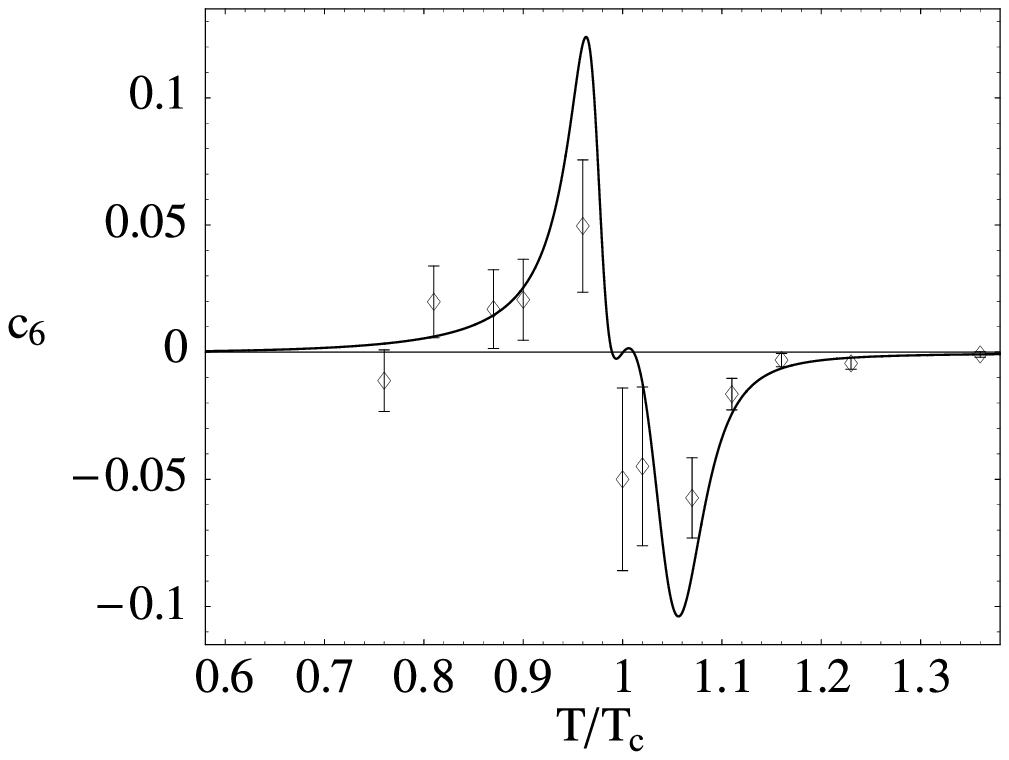}
\end{minipage}
\hfill
\begin{minipage}[t]{.48\textwidth}
\includegraphics*[width=\textwidth]{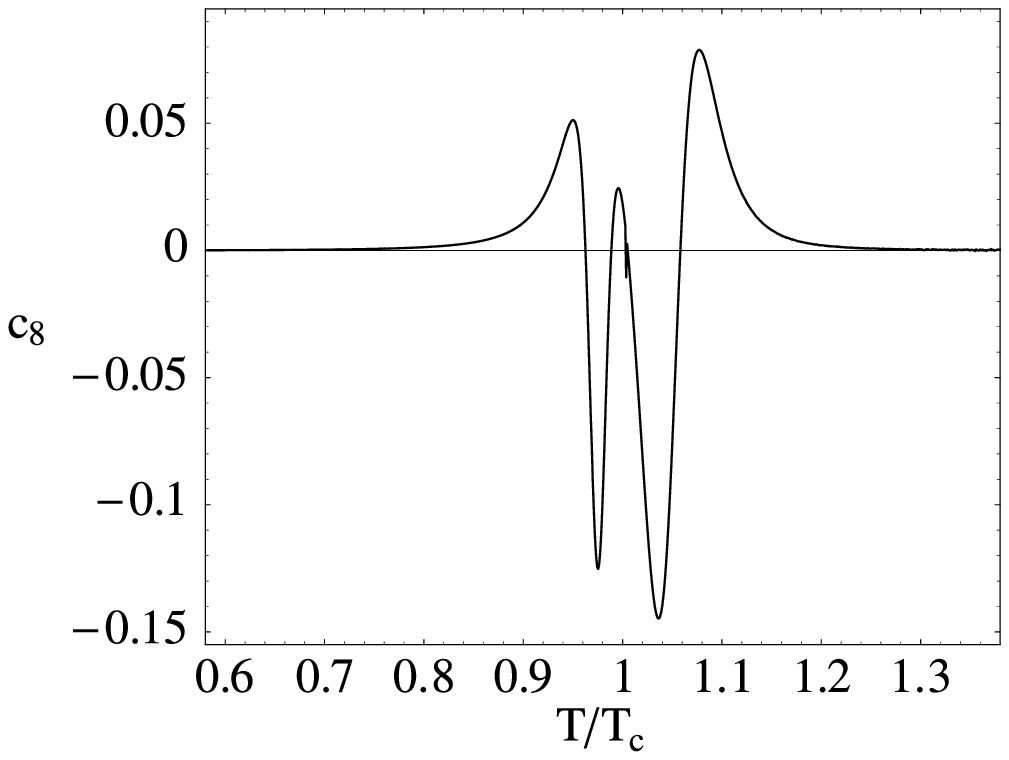}
\end{minipage}
\caption{
\footnotesize 
Second, fourth, sixth and eighth moments of the pressure difference with respect to the 
chemical potential, plotted as functions of the temperature. (Note that the temperature scales of the upper and lower graphs are different.) We compare to lattice data (diamonds with errorbars) taken from Ref.~\cite{Allton:2005gk} 
\label{fig:moments}}
\end{figure}

Lattice results at finite quark chemical potential
are obtained as Taylor expansions of the thermodynamical quantities in the parameter 
$\mu/T$ around zero chemical potential. Here we perform 
the same kind of expansion in the PNJL model and compare with Taylor coefficients deduced from lattice data. Examples are the coefficients in the expansion of the pressure $p = -\Omega$:
\bea
\frac{p(T,\mu)}{T^4}=\left.\sum_{n=0}^{\infty}c_n(T)
\left(\frac{\mu}{T}\right)^n~~~~~~\mathrm{with}~~~~~~c_n(T)
=\frac{1}{n!}\frac{\partial^n (p(T,\mu)/T^4)}{\partial(\mu/T)^n}\right |_{\mu=0}
\eea
and even $n$. Specifically:
\begin{align}
 & \;c_2 = \left.\frac{1}{2}\,\frac{\partial^2 (p/T^4)}{\partial(\mu/T)^2}\right|_{\mu=0}, &
 & \;\,c_4 = \left.\frac{1}{24}\,\frac{\partial^4 (p/T^4)}{\partial(\mu/T)^4}\right|_{\mu=0}, \nonumber\\
 & c_6 = \left.\frac{1}{720}\,\frac{\partial^6 (p/T^4)}{\partial(\mu/T)^6}\right|_{\mu=0}, &
 & c_8 = \left.\frac{1}{40320}\,\frac{\partial^8 (p/T^4)}{\partial(\mu/T)^8}\right|_{\mu=0}.
\end{align}
Results for these coefficients are shown in 
Fig.~\ref{fig:moments}. We notice in particular the remarkably good agreement between the calculated "susceptibility" $c_4$ and the lattice data. This quantity has recently been computed in Ref.~\cite{Ghosh:2006qh} using the previous version of our PNJL model, Ref.~\cite{ratti}, with a less satisfactory outcome. 
Now, with the improved effective potential ${\cal U}$ as described in Eq.(\ref{u1}), the agreement is significantly better. 
We suspect that previous calculations did not approach the Stefan-Boltzmann limit in an acceptable way \cite{Mukherjee:2006hq}, due to the large gap between $\Phi$ and $\Phi^*$, that persisted up to rather high temperatures. As discussed above this large split occurs when not properly keeping all constraints on $\phi_3$ and $\phi_8$ under control.

\subsection{Phase diagram}

\begin{figure}
\begin{minipage}[t]{.46\textwidth}
\includegraphics*[width=\textwidth, clip=true, trim=0 0 0 0 ]{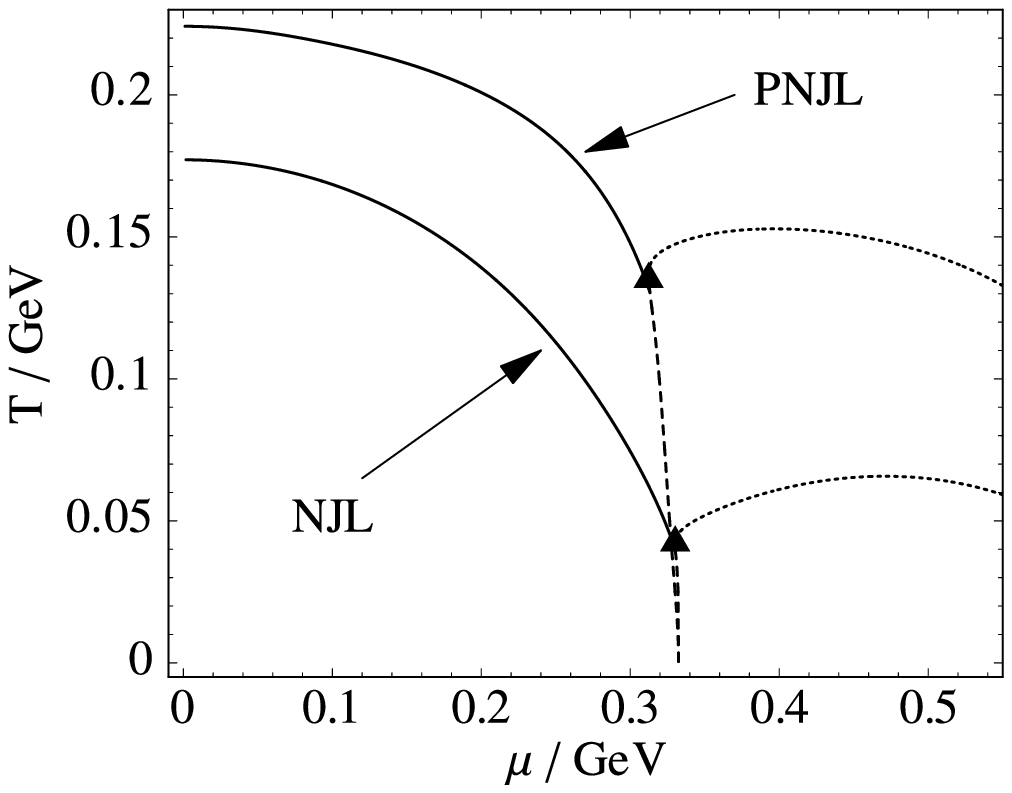}
\end{minipage}
\hfill
\begin{minipage}[t]{.46\textwidth}
\includegraphics*[width=\textwidth, clip=true, trim=0 0 0 0 ]{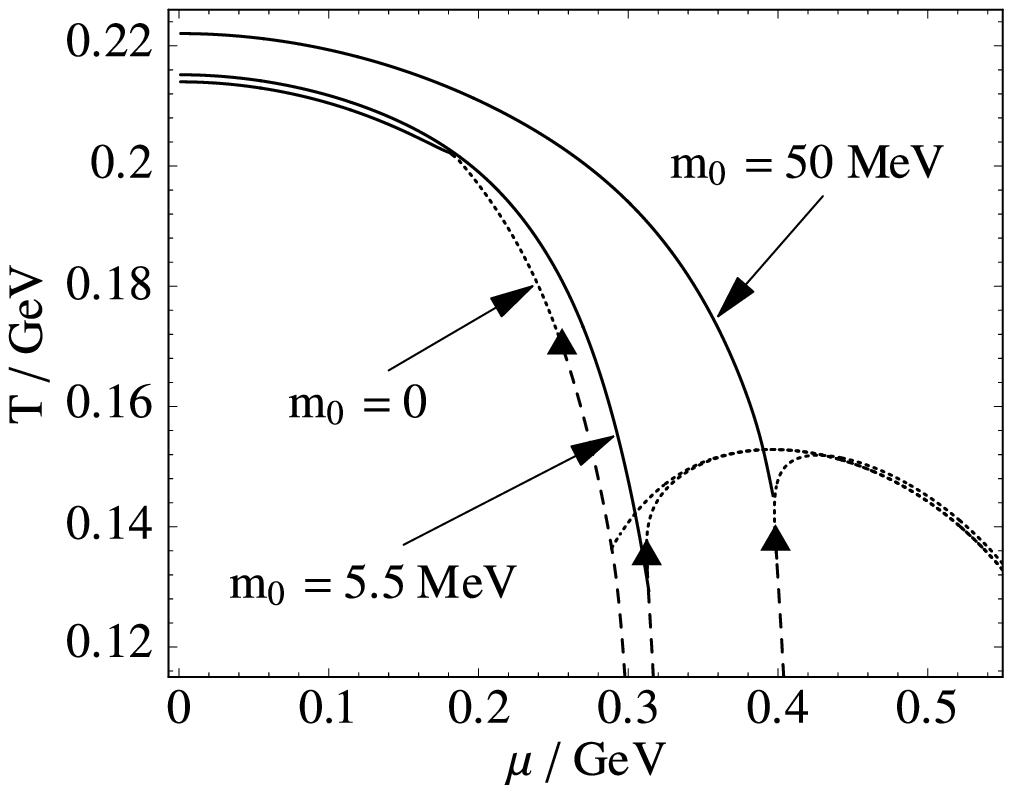}
\end{minipage}
\\
\begin{minipage}[t]{\textwidth}
\centering
\includegraphics*[width=.46\textwidth, clip=true, trim=0 0 0 0 ]{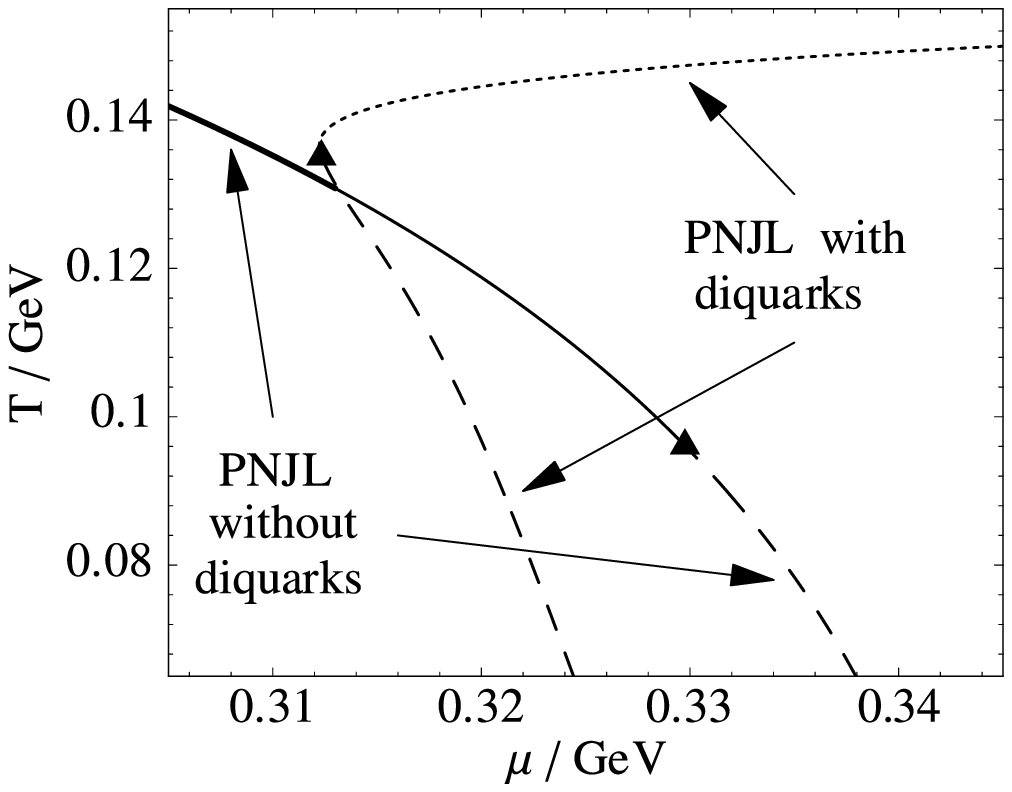}
\end{minipage}
\caption{
\footnotesize 
Upper left panel: comparison of the phase diagrams of NJL and PNJL model. The cross-over of the chiral condensate is drawn solid, first-order lines are dashed and second order lines dotted.
Upper right panel: comparison of the phase diagram at different current quark masses with inclusion of diquark degrees of freedom. (Note the scale on the temperature axis.)
Lower panel: comparison of the PNJL model with and without inclusion of diquarks.
\label{fig:phasediagram}}
\end{figure}

We now turn to the phase diagram in the $(T,\mu)$ plane as derived from this updated version
of the PNJL model.
The left panel of Fig.~\ref{fig:phasediagram} shows the phase diagrams in the $(T,\mu)$-plane computed using the PNJL model in comparison with the NJL model (the limiting case in which $\Phi\equiv 1$). Of particular interest is the location of the critical endpoint at which the chiral and deconfinement crossover transitions at lower $\mu$ turn into a first-order phase transition above some critical $\mu$. 
The crossover is not a phase transition. Therefore there exist several ways to locate the position of a crossover transition. In the present calculations the crossover line is determined using the order parameters in the chiral limit (the chiral condensate) and the pure gauge theory (the Polyakov loop) respectively. Since these order parameters show their strongest changes as functions of temperature along the crossover transition lines, we determine their position by local maxima of $\dd\sigma/\dd T$ and $\dd \Phi/\dd T$ \footnote{Other frequently used and closely related criteria for the definition of crossover transition lines involve chiral or Polyakov loop susceptibilities. This does not lead to any significant differences for the phase diagram in comparison with the method applied here.}.

The crossover transition lines fixed by either the susceptibilities of $\sigma$ and $\Phi$ or by maximal changes with temperature, i.e. zeros of $\dd^2\sigma/\dd T^2$ or $\dd^2\Phi/\dd T^2$, do coincide with the critical point for our PNJL model in the absence of diquarks (see lower panel of Fig.~\ref{fig:phasediagram}). This is a consequence of the divergences in these quantities at the critical point. However, when including diquarks, a coincidence of critical point and crossover transition line is not guaranteed.

One finds that the critical endpoint depends sensitively on the degrees of freedom involved.
From its position in the restricted NJL case (see also \cite{Buballa}) this point is shifted to higher $T$ by both, the effective Polyakov loop potential, and by the presence of diquark degrees of freedom. Near the critical endpoint not including diquarks, $\frac{\dd \sigma}{\dd T}$ diverges together with the chiral susceptibility. This extreme behaviour is not observed in the case with inclusion of diquarks. The region where this critical behaviour would appear is now already located in the diquark dominated phase.

Thus there is a qualitative difference of the critical endpoints in these two compared cases: not including diquarks the critical endpoint lies on top of the merging chiral and deconfinement crossover transition lines, while in the case including diquarks the critical endpoint is shifted away from this line. The critical endpoint now lies on the second order transition line bordering the diquark dominated phase (see lower panel of Fig.~\ref{fig:phasediagram}).  I.\,e. the endpoint is not at the junction of all three transition lines and therefore is not a tri-critical point but still a critical point.

Next we use the PNJL model including diquark degrees of freedom to study the dependence of the position of the critical endpoint on the bare (current) quark mass. The upper right panel of Fig.~\ref{fig:phasediagram} shows phase diagrams in the chiral limit, for current quark masses $m_0 = 5.5\,\mathrm{MeV}$ and $m_0 = 50\,\mathrm{MeV}$. The change of the critical endpoint with varying quark mass mainly reflects the dependence of the critical chemical potential on the quark mass. The presence of the diquark dominated phase appears to stabilize the temperature of the critical endpoint at rather high values. 

\begin{figure}
\begin{center}
\includegraphics*[width=.46\textwidth]{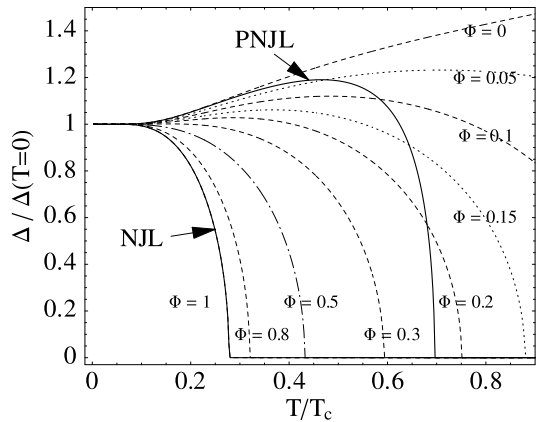}
\caption{
\footnotesize Dependence of the gap $\Delta$ on the presence of the Polyakov loop. The solid lines are the solutions to the self consistency equations of the NJL and the PNJL model at $\mu = 0.4\,\mathrm{GeV}$. The dashed lines are obtained by enforcing fixed values for the Polyakov loop. Note that the PNJL model with the Polyakov loop fixed at $\Phi = 1$ (deconfinement) coincides with the self consistent solution of the NJL model.
\label{fig:gap}}
\end{center}
\end{figure}

Generally, the PNJL model generates the critical endpoint at a temperature which is significantly higher than the one found with the standard NJL model, i. e. ignoring Polyakov loop dynamics. The reason is that the diquark phase as well as the chiral phase is stabilized by the confinement emulation via the effective Polyakov loop potential. The size of the gap $\Delta$ is strongly influenced by the Polyakov loop. The detailed dependence of the gap on the Polyakov loop is displayed in Fig.~\ref{fig:gap}. The systematics of this effect becomes evident when the Polyakov loop is held at fixed values and varied. The gap resulting from this calculation is then compared to the gap in the PNJL model (with self-consistent determination of $\Phi$) and in the NJL model. The case where the Polyakov loop is fixed to $\Phi=1$ (i.\,e. complete deconfinement) coincides with the NJL calculation. 

The presence of the Polyakov loop restricts the phase space available for quarks in the vicinity of their Fermi surface where Cooper pair condensation takes place. Hence a higher temperature is effectively required to break the pairs. This is the primary reason for
the difference in behavior of the gap $\Delta$ when comparing NJL and PNJL results in Fig.~\ref{fig:gap}.  

\subsection{Speed of sound}

\begin{figure}
\begin{center}
\begin{minipage}[t]{.47\textwidth}
\includegraphics*[height=165pt]{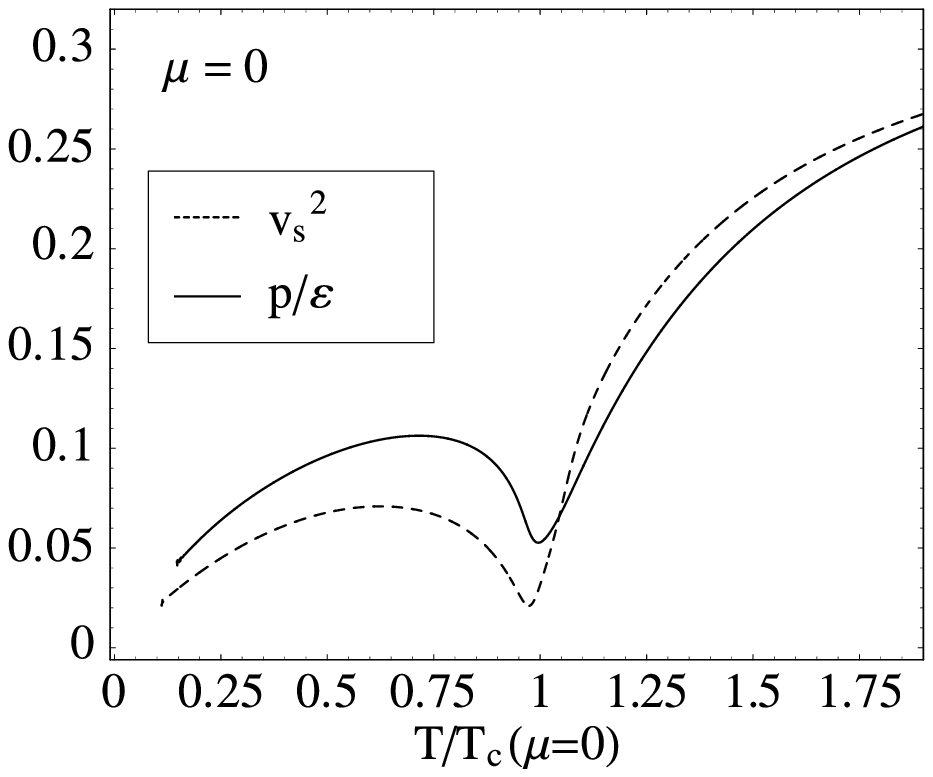}
\end{minipage}
\hfill
\begin{minipage}[t]{.25\textwidth}
\includegraphics*[height=165pt, clip=true, trim=0 9.5 0 7.5 ]{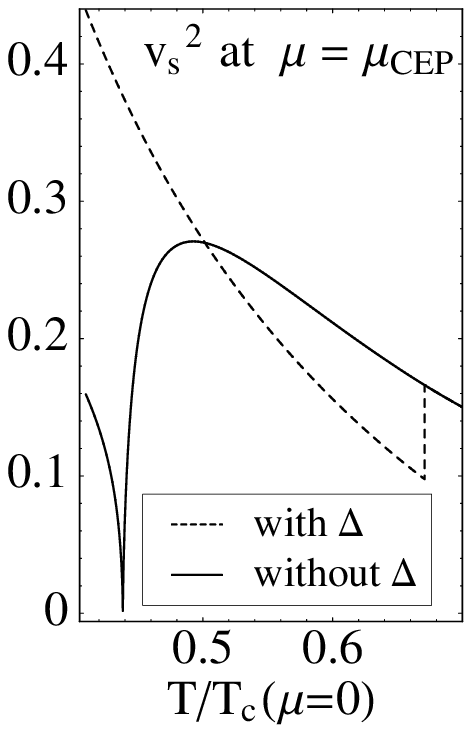}
\end{minipage}
\hfill
\begin{minipage}[t]{.25\textwidth}
\includegraphics*[height=165pt, clip=true, trim=0 9.5 0 7.5 ]{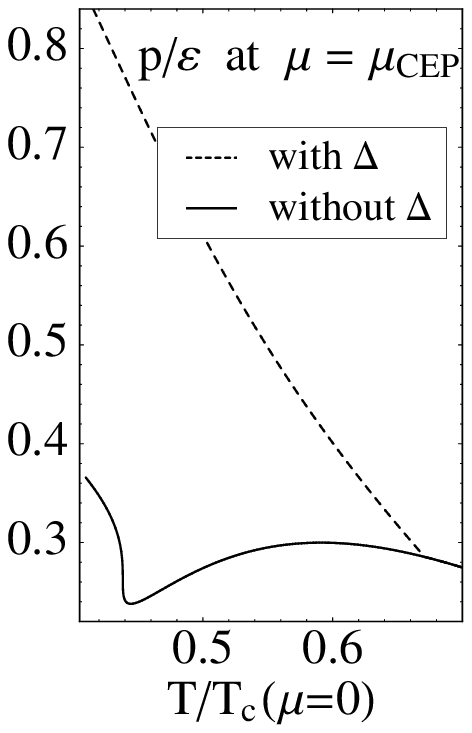}
\end{minipage}
\caption{
\footnotesize
Left panel: Squared speed of sound (dashed) and ratio $\frac{p}{\varepsilon}$ (solid), and their dependence on the temperature at vanishing chemical potential.
The two panels to the right: Studies of the sound velocity in the vicinity of the critical endpoint in the PNJL model without diquarks (at chemical potential $\mu = \mu_{\mathrm{CEP}}$). The PNJL model without diquarks is drawn with solid lines, the PNJL model with diquarks with dashed lines.
Center panel: square of the velocity of sound.
Right panel: pressure over energy density $\frac{p}{\varepsilon}$.
\label{fig:speed_of_sound}}
\end{center}
\end{figure}

The velocity of sound $v_s$, determined by
\beq
 v_s^2 = -\left. \frac{1}{C_V} \frac{\, \partial\Omega \,}{\partial T} \right\vert_V
\label{sound}
\eeq
with the specific heat $C_V = -T(\partial^2\Omega/\partial T^2)_V$,
shows a pronounced dip near the chiral and the deconfinement transition. This local minimum of the speed of sound becomes deeper in the vicinity of the critical endpoint of the first-order phase transition line, separating the chiral phase at low chemical potential ($\mu \lesssim 1.5\,T_c $) from the diquark phase at high chemical potential ($\mu \gtrsim  1.5\,T_c $). When neglecting diquark degrees of freedom the speed of sound vanishes at the critical endpoint (solid curve in the central panel of Fig.~\ref{fig:speed_of_sound}).

Correspondingly, the specific heat diverges at this point.
When diquark degrees of freedom are included in the calculation the critical endpoint is shifted such that the region of vanishing speed of sound would already be placed within the diquark dominated phase. This is why a vanishing speed of sound is not observed in the model including explicitly diquark degrees freedom (dashed curve in the central panel of Fig.~\ref{fig:speed_of_sound}). The discontinuity at higher temperatures is generated by the second order phase transition separating the diquark dominated phase from the high temperature phase. Above this transition the two versions of the PNJL model (with and without explicit diquarks) become equivalent. 

\section{Concluding remarks and outlook}

We have pointed out that an updated version of the PNJL model over and beyond the one used in \cite{ratti,Ghosh:2006qh} leads to significantly better agreement with lattice data \cite{Allton:2005gk,Kaczmarek:2005}, especially when extrapolating to finite chemical potential $\mu$. The combination of only two principal ingredients: chiral symmetry restoration and an effective potential ansatz for the confinement order parameter, appears to be sufficient to reproduce the available full QCD lattice computations to an astonishingly high accuracy, at least for temperatures $T$ up to about $2T_c$.
The improvements shown in this paper in comparison with previous results \cite{ratti,Ghosh:2006qh} originate in a better representation of the Polyakov loop part of the PNJL model. Taking into account the proper $\mathrm{SU(3)}$ constraints is crucial for an effective description of the thermodynamical implications of confinement.

Incorporating explicit diquark degrees of freedom influences the position and the nature of the critical endpoint in the $(T,\mu)$ phase diagram.
The critical endpoint in the presence of diquarks is the connecting point between the chiral crossover transition line and the second order transition bordering the diquark dominated phase, while in the absence of diquarks it is the junction point of the chiral and deconfinement crossover transition. The critical point in the PNJL model with diquarks turns out not to coincide with the critical (diverging) behaviour of susceptibilities related to the chiral condensate and the Polyakov loop. 

Further developments now aim for an extension of the present framework to $N_f = 3$ in order to explore the rich structure of colour superconducting (diquark) phases with three quark flavours and the additional effects of Polyakov loop dynamics.


\begin{thebibliography}{999}

\bibitem{phil}
P. de Forcrand and O. Philipsen, Nucl. Phys. B {\bf 642}, 290 (2002);
Nucl. Phys. B {\bf 673}, 170 (2003).

\bibitem{katz}
Z.~Fodor and S.~D. Katz,
\newblock JHEP {\bf 0203}, 014 (2002); Z.~Fodor, S.~D. Katz, and K.~K. Szabo,
Phys. Lett. B {\bf 568}, 73 (2003).

\bibitem{Allton1}
C.~R. Allton {\em et~al.},
\newblock Phys. Rev. D {\bf 66}, 074507 (2002); Phys. Rev. D {\bf 68}, 014507 (2003).

\bibitem{Allton:2005gk}
C.~R. Allton {\em et~al.},
\newblock Phys. Rev. D {\bf 71}, 054508 (2005).

\bibitem{Boyd}
G.~Boyd {\em et~al.},
\newblock Nucl. Phys. B {\bf 469}, 419 (1996).

\bibitem{Kaczmarek:2002mc}
O.~Kaczmarek, F.~Karsch, P.~Petreczky, and F.~Zantow,
\newblock Phys. Lett. B {\bf 543}, 41 (2002).

\bibitem{Kaczmarek:2005}
O.~Kaczmarek and F.~Zantow,
\newblock Phys. Rev. D {\bf 71}, 114510 (2005).

\bibitem{Fukushima:2003fw}
K.~Fukushima,
\newblock Phys. Lett. B {\bf 591}, 277 (2004).

\bibitem{Meisinger:1995}
  P.~N.~Meisinger and M.~C.~Ogilvie,
\newblock  Nucl.\ Phys.\ Proc.\ Suppl.\  {\bf 47}, 519 (1996);
  Phys.\ Lett.\ B {\bf 379}, 163 (1996). 

\bibitem{ratti}
C. Ratti, M. A. Thaler and W. Weise, Phys. Rev. D {\bf 73}, 014019 (2006);\\ 
C. Ratti, M. A. Thaler and W. Weise, nucl-th/0604025.

\bibitem{Ratti:2006wg}
  C.~Ratti, S. R\"o{\ss}ner, M. A. Thaler and W. Weise, Eur. Phys. J. C (2006), in print,
  arXiv:hep-ph/0609218.

\bibitem{Ghosh:2006qh}
S.~K.~Ghosh, T.~K.~Mukherjee, M.~G.~Mustafa and R.~Ray,
Phys.\ Rev.\ D {\bf 73}, 114007 (2006).

\bibitem{Zhang:2006gu}
  Z.~Zhang and Y.~X.~Liu,
  arXiv:hep-ph/0610221.

\bibitem{dpz05}
A. Dumitru, R. D. Pisarski and D. Zschiesche,
\newblock Phys. Rev. D {\bf 72}, 065008 (2005).

\bibitem{simon1}
S. R\"o\ss ner, Diploma Thesis, Technical University of Munich (2006).

\bibitem{simon2}
S. R\"o\ss ner, C. Ratti, W. Weise, in preparation.

\bibitem{Fukushima:2006uv}
  K.~Fukushima and Y.~Hidaka,
  arXiv:hep-ph/0610323.

  \bibitem{Boyd:1995cw}
  G.~Boyd, S.~Gupta, F.~Karsch, E.~Laermann, B.~Petersson and K.~Redlich,
\newblock Phys.\ Lett.\ B {\bf 349}, 170 (1995).

\bibitem{Mukherjee:2006hq}
  S.~Mukherjee, M.~G.~Mustafa and R.~Ray,
  arXiv:hep-ph/0609249.

\bibitem{Buballa}
M.~Buballa,
Phys. Reports {\bf 407}, 205 (2005).


\end{thebibliography}
\end{document}